\documentclass[sts]{imsart}

\RequirePackage{dsfont, amsthm,amsmath,amsfonts,amssymb}
\RequirePackage[numbers,sort&compress]{natbib}
\RequirePackage[colorlinks,citecolor=blue,urlcolor=blue]{hyperref}
\RequirePackage{graphicx}

\startlocaldefs
\theoremstyle{plain}

\theoremstyle{definition}

\endlocaldefs
\usepackage{bbm}
\begin{document}

\begin{frontmatter}
\title{Longitudinal Outcomes Truncated by Death: Causal Estimands and Bayesian Estimators}
\runtitle{Longitudinal Outcomes Truncated by Death: Causal Estimands and Bayesian Estimators}

\begin{aug}
\author[A]{\fnms{Juliette}~\snm{Ortholand}\thanks{\textbf{Corresponding author}}\ead[label=e1]{first@somewhere.com}}
\author[B]{\fnms{Young}~\snm{Lee}\ead[label=e2]{second@somewhere.com}}
\author[C]{\fnms{Marie-Abèle}~\snm{Bind}\ead[label=e3]{third@somewhere.com}}

\address[A]{Juliette Ortholand, Medical Informatics Department,
Amsterdam UMC, Amsterdam, the Netherlands.}

\address[B]{Young Lee, Singapore University of Technology and Design.}

\address[C]{Marie-Abèle Bind, Biostatistics Center at Massachusetts General Hospital and Department of Medicine, Harvard Medical School, Boston, USA.}

\end{aug}

\begin{abstract}
In randomized controlled trials with longitudinal outcomes, death before the end of follow-up poses a fundamental challenge: after death, the outcome is no longer a real-valued measurement. This complicates the definition and interpretation of causal estimands, particularly when treatment may affect both survival and longitudinal outcomes.

We review existing estimands for longitudinal outcomes truncated by death and clarify the assumptions required for their identification and estimation. We show that these estimands fall into two broad classes, distinguished by whether they require additional assumptions to compare longitudinal outcomes beyond death. Such assumptions may be inappropriate in chronic diseases, either because i) death and longitudinal outcomes are driven by the same underlying biological process or ii) the relative desirability of survival with poor function versus death may depend on individual preferences.

We compare the behavior of the estimands in a simulation study using Bayesian estimators and illustrate their use with data from a randomized controlled trial in amyotrophic lateral sclerosis. We argue that, in the presence of death truncation, pairing the survivor average causal effect with the restricted mean survival time estimand provides an interpretable characterization of treatment effects on longitudinal and survival outcomes.
\end{abstract}


\end{frontmatter}

\section{Introduction}

Scores or biomarkers, typically measured longitudinally, are key outcomes in clinical analysis and are increasingly regarded as essential even in settings with high mortality \cite{thomassen_role_2024}. In randomized controlled trials (RCTs) with longitudinal outcomes, death before the end of follow-up poses a fundamental challenge: after death, the outcome is no longer a real-valued measurement \cite{frangakis_principal_2002}. This complicates the definition and interpretation of causal estimands, particularly when treatment may affect both survival and longitudinal outcomes.

Some causal estimands have been proposed to address this issue. The survivor average causal effect enables focusing on the treatment effect for patients that would have survived in both arms \cite{frangakis_principal_2002}. The Wilcoxon-Mann-Whitney estimand combines survival with the longitudinal outcome to get an overview of the treatment effect on the whole population \cite{fay_causal_2018}. Yet, their use in practice remains limited \cite{thomassen_role_2024}.
In parallel, estimators have been developed to tackle this problem, notably based on landmarking \cite{anderson_analysis_1983} and joint modeling approaches \cite{rizopoulos_joint_2012}. However, these methods are typically not accompanied by an explicit formulation of the causal estimand they estimate. This contrasts with the recommendations of the European Medicines Agency, which emphasized the central role of estimands in the ICH E9 addendum \cite{european_medicines_agency_ema_e9_2020} and recent scientific work advocating a shift in focus away from modeling alone \cite{carlinmoreno-betancurUsesAbusesRegression2024}. Consistent with this, a recent review of methods for handling death in RCTs summarized several analytical strategies, yet without analysing them in terms of potential outcomes \cite{thomassen_role_2024}.

RCTs conducted on Amyotrophic lateral sclerosis (ALS) are a concrete illustration of this problem. ALS is a neurodegenerative disorder characterized by progressive loss of motor function and death from respiratory failure \cite{talbott_epidemiology_2016}. The median survival time from disease onset is approximately three to four years \cite{talbott_epidemiology_2016}. The U.S. Food and Drug Administration requires ALS RCTs to assess treatment effects on both overall survival and longitudinal outcomes capturing functional abilities \cite{FDA_adjusting_2023}. Individuals’ functional decline, evaluated through functional ability in daily activities, is most commonly assessed using the revised ALS functional rating scale (ALSFRS-R) \cite{rooney_what_2017}. The motivating example in this article concerns the estimation of a treatment effect using ALSFRS-R. Accordingly, several ALS-specific outcomes were designed for ALS RCTs \cite{quintana_design_2023,berry_combined_2013}, and the two causal estimands previously mentioned were used \cite{van_eijk_composite_2022, rubinCausalInferencePotential2006}.

In this context, our contribution is twofold. First, in Section \ref{sec:estimand}, we review several existing estimands, examining how their construction addresses the fact that, following death, the longitudinal outcome is no longer a real-valued measurement \cite{frangakis_principal_2002}.
Second, we assess their interpretability and practicality. For each selected estimand, we make explicit the assumptions required for the estimation of unbiased estimates (section \ref{sec:estimation}). We then illustrate their statistical behavior with Bayesian estimators on simulated data (section \ref{sec:simulation}). Finally, we apply them using data from an RCT in ALS (section \ref{sec:application}). 

\section{Notations and Science Table}\label{sec:notations}

We consider a population of $N$ patients indexed by $i=1,\dots, N$, each assigned a binary treatment $W_i\in\{0,1\}$, where $W_i=1$ denotes the active treatment and $W_i=0$ the control treatment. At each visit time $t$, individuals have two longitudinal potential outcomes $Y_{i,t}(0), Y_{i,t}(1)$. Only one of them is observed $Y^{\mathrm{obs}}_{i,t} = Y_{i,t}(W_i)$ and the other is missing $Y^{\mathrm{mis}}_{i,t} = Y_{i,t}(1-W_i)$. As for the ALSFRS-R score, we consider lower scores indicating more severe disease.  Time to death is denoted with two potential outcomes. The potential times of death or administrative censoring at $t$ are denoted by $T_{i,t}(0)$ and $T_{i,t}(1)$. We associate them with potential event indicators $(D_{i,t}(0),D_{i,t}(1)$), such that $D_{i,t}(W)=1$ if death occurs by time $t$, and $D_{i,t}(W)=0$ otherwise. Together $(T_{i,t}(W), D_{i,t}(W))$ are referred to as the time-to-event potential outcomes. Similarly, $(T_{i,t}^{\mathrm{obs}}, D_{i,t}^{\mathrm{obs}})$ denote the observed outcomes and $(T_{i,t}^{\mathrm{mis}}, D_{i,t}^{\mathrm{mis}})$ the missing ones. After death, the longitudinal outcome is no longer a real-valued measurement and can be denoted by $*$ \cite{frangakis_principal_2002}. Baseline covariates measured before randomization are denoted $X_i$. The science table (Table \ref{tab:science_table}) illustrates longitudinal and time-to-event outcomes for a small set of hypothetical individuals. 

Real numbers and $*$ cannot be compared, and this complicates the definition of an estimand. Yet, each individual can be classified into one of four strata \cite{frangakis_principal_2002}, denoted by $G_{i,t}$: 
\begin{itemize}
    \item Always-survivor ($\mathrm{LL}_t$): lives regardless of treatment assignment,
    \item Protected ($\mathrm{DL}_t$): dies under control treatment, but lives under active treatment,
    \item Harmed ($\mathrm{LD}_t$): lives under control treatment, but dies under active treatment,
    \item Never-survivor ($\mathrm{DD}_t$): dies regardless of treatment assignment.
\end{itemize}

\begin{table*}
    \centering
    \begin{tabular}{cccccccccccccccc}
\hline
Individuals & Time & Treatment & Baseline covariates &\multicolumn{3}{c}{Longitudinal outcome}& \multicolumn{3}{c}{Indicator of death} & \multicolumn{3}{c}{Time of death}\\
 $i$ & $t$ (months) & $W_i$ & $X_i$ &$Y_{i,t}(0)$ & $Y_{i,t}(1)$ & $Y^{\mathrm{obs}}_{i,t}$& $D_{i,t}(0)$ & $D_{i,t}(1)$ & $D^{\mathrm{obs}}_{i,t}$ & $T_{i,t}(0)$ & $T_{i,t}(1)$ & $T^{\mathrm{obs}}_{i,t}$\\
\hline
1 & 6 & 0 &$X_1$ &-2 & ? & -2 &  0 & ? & 0& 6 &? &6 \\
1 & 18 & 0 &$X_1$ &-4 & ? & -4 &  0 & ? & 0& 18 &? &18 \\
2 & 6 & 0 & $X_2$ & -3 & ? & -3 &   0 & ? & 0 & 6 & ? & 6 \\
2 & 18 & 0 & $X_2$ & -6 & ? & -6 &   0 & ? & 0 & 18 & ? & 18 \\
3 & 6 & 1& $X_3$ & ? & -3 & -3 &   ? & 0 & 0 & ? & 6 & 6 \\
3 & 18 & 1& $X_3$ & ? & -5 & -5 &   ? & 0 & 0 & ? & 18 & 18 \\
4 & 6 & 1 &$X_4$ & ? & * & * &  ? & 0 & 0 & ? & 6 & 6\\
4 & 18 & 1 &$X_4$ & ? & * & * &  ? & 1 & 1 & ? & 8 & 8\\
\hline
    \end{tabular}
    \caption{Science Table\\
    \underline{Legend:} ?: missing value, *: does not belong to the outcome space of real numbers}
    \label{tab:science_table}
\end{table*}

\section{Causal Estimands} \label{sec:estimand}

Individuals enrolled in RCTs are rarely representative of the general population. Accordingly, we focus on finite-sample estimands and do not consider generalization to a superpopulation \cite{colnetCausalInferenceMethods2023}. In this section, we review different causal estimands. We first consider estimands for longitudinal outcomes not truncated by death and for time-to-event outcomes. We then present two classes of estimands that require assumptions to compare $*$ and real numbers. We conclude with an estimand that does not require such assumptions.
Causal estimands are defined by comparing the potential outcomes under the two treatment conditions, for each individual, at the same time point $t$ since randomization \cite{imbens_causal_2015}. Throughout this section, to make this definition explicit, we use potential outcome notation $(Y_{i,t}(0), Y_{i,t}(1))$, rather than observed/missing outcome notation $(Y^{\mathrm{obs}}_{i,t}, Y^{\mathrm{mis}}_{i,t})$.

\subsection{Classic Causal Estimands}

In the absence of truncation due to death, the average causal effect (ACE) can be used to define the treatment effect on the longitudinal outcome. It is the average over individuals, of the difference of the two longitudinal potential outcomes at time $t$:
\begin{eqnarray*}
\tau^{\mathrm{ACE}}_{\mathrm{fs}}(t)
&=
\frac{1}{N}\displaystyle\sum_{i=1}^{N}\left\{Y_{i,t}(1)-Y_{i,t}(0)\right\}
\end{eqnarray*}
%
In time-to-event analysis, two fundamental quantities are the survival ($S_{i,t}(W)$) and the hazard ($\Lambda_{i,t}(W)$). The survival is the probability of remaining event-free beyond time $t$ (Equation \ref{eq:survival_function}).  The hazard is the instantaneous event rate at time $t$, conditional on survival up to time $t$ (Equation \ref{eq:hazard_function}). 
\begin{eqnarray}
S_{i,t}(W)
=&
\Pr\!\left(T_i(W) > t\right) \label{eq:survival_function}\\
\Lambda_{i,t}(W)
=&
\lim_{dt \to 0}
\frac{\Pr\!\left(t \leq T_i(W) < t + dt \,\middle|\, T_i(W) \geq t\right)}{dt} \label{eq:hazard_function}
\end{eqnarray}
These quantities form the foundation for defining causal estimands in time-to-event settings.
One well-known hazard-based estimand is the hazard ratio: 
\begin{equation*}
    \tau^{HR}_{\mathrm{fs}}(t) = \frac{1}{N} \displaystyle\sum_{i=1}^N \left\{\frac{\Lambda_{i,t}(1)}{\Lambda_{i,t}(0)} \right\}
\end{equation*}
However, to be used, the hazard requires that the individual is alive in both arms at time $t$, which complicates its usability \cite{fay_causal_2024}. This complexity has led to an increasing popularity of a survival-based estimand: the restricted mean survival time (RMST) \cite{hanRestrictedMeanSurvival2022, zhaoRestrictedMeanSurvival2016, zhao_utilizing_2012, roystonUseRestrictedMean2011, karrison_use_1997}. It can be interpreted as "the average survival time or life expectancy during a defined time period ranging from time 0 to a specific follow-up time point" \cite{hanRestrictedMeanSurvival2022}.
\begin{eqnarray}
    \nonumber \tau^{RMST}_{\mathrm{fs}}(t) =& \frac{1}{N} \displaystyle\sum_{i=1}^N \left\{ \int^t_0 S_{i,u}(1) -  S_{i,u}(0) du \right\} \\
    =& \frac{1}{N} \displaystyle\sum_{i=1}^N \left\{ T_{i,t}(1) - T_{i,t}(0) \right\} \label{eq:estimand_rmst}
\end{eqnarray}
Due to its clear causal and clinical interpretation, the RMST is adopted as the causal estimand to study death in this article.

\subsection{Hypothetical estimands}

An option to overcome the problem of truncation by death is to replace * by the longitudinal outcome value that individuals would have had if they were not dead. For this approach to have a causal interpretation, one should assume the existence of a hypothetical intervention that prevents death while leaving the longitudinal trajectory unchanged \cite{olarte_parra_hypothetical_2023, olarteparra_estimating_2025}. Prominent examples of such hypothetical estimands include the controlled direct effect, originally introduced by Robins and Greenland \cite{robins_identifiability_1992}.
Yet, in most chronic diseases, the same underlying biological process drives both functional decline and death. Thus, although such estimands may be meaningful in certain settings, the assumption that death can be intervened upon without affecting the longitudinal outcome is highly implausible in our context. Consequently, these estimands generally lack a coherent clinical interpretation and are not well-suited to the research questions considered in this article.

\subsection{Assuming an Order: Composite Causal Estimand}

Another option to overcome the problem of truncation by death is to use time-to-event potential outcomes. Yet, such an approach requires adding an assumption to compare death and longitudinal outcomes. A common choice is to regard death as the worst possible outcome \cite{pocock_win_2012, van_eijk_composite_2022}. However, this is a strong assumption, particularly in chronic and fatal diseases, where individuals may reasonably prefer a shorter survival with better quality of life, to a longer survival with severe disability. 
Under this assumption, a composite potential outcome, that represents a general notion of health, can be defined as a three-dimensional vector comprising: the longitudinal, the time of event, and the event indicator potential outcomes.
\begin{eqnarray*}
Z_{i,t}(W)
&=
\Bigl(Y_{i,t}(W),\, T_{i,t}(W),\, D_{i,t}(W)\Bigr)
\end{eqnarray*}
If an individual belongs to the always-survivor stratum, the longitudinal potential outcome can be used to compare composite potential outcomes. If the individual is dead under one treatment, one of the longitudinal potential outcomes is *, thus the longitudinal potential outcome cannot be used anymore. Yet, we can use the time-to-event potential outcomes to compare the composite potential outcomes. Following this logic, we can define a one-dimensional treatment effect as the difference of the three-dimensional composite potential outcomes, for each stratum to which the individual might belong: protected, harmed, always-survivors, and never-survivors (see section \ref{sec:notations}). Assuming death is the worst outcome, individuals from the protected stratum have the best treatment effect, and individuals from the harmed stratum the worst. To avoid combining quantities with incompatible units, the treatment effect is defined as taking infinite values. The difference of composite potential outcomes is thus set to $+\infty$ for individuals in the protected stratum and $-\infty$ for individuals in the harmed stratum. 

For individuals that belong to the always-survivor stratum, we can use the longitudinal potential outcomes, so that the difference of composite potential outcomes coincides with that of the longitudinal potential outcomes: 
\begin{align}
    Z_{i,t}(1)-Z_{i,t}(0) = Y_{i,t}(1)-Y_{i,t}(0) \text{ if } G_i = LL_t \label{eq:comp_long_equal}
\end{align} 
Finally, for individuals who belong to the never-survivor stratum, the difference in time of event potential outcomes can be used to define an order. Yet, to avoid having treatment effects in different units, the difference is set to $+\infty$ if survival time is higher under active treatment and $-\infty$ otherwise. The difference is set to zero if an individual dies at the same time under both treatments. Table~\ref{tab:composit_outcome} summarizes this ordering, with the value of the difference of the composite potential outcomes.

\begin{table*}
\begin{center}
\begin{tabular}{ccccrclc}
\hline
 Order & Stratum &&  Additional &\multicolumn{3}{c}{Composite potential outcome $Z_{i,t}(W)$}  &        Metric               \\ 
  &   $G_{i,t}$& & order condition &\multicolumn{3}{c}{$\Big(Y_{i,t}(W),T_{i,t}(W), D_{i,t}(W)\Big)$}  &                $Z_{i,t}(1) - Z_{i,t}(0)$        \\ 
 &       \multicolumn{3}{c}{}  &$W =0$ &&  $W =1$  &                    \\ 
 \hline
 $+++$                   & $ \mathrm{\textcolor{red}{D}L}_t$ &   &     &\textcolor{red}{$\Big(*, T_{i,t}(0), 1\Big)$}  &$<$& $\Big(Y_{i,t}(1),T_{i,t}(1), 0\Big)$ & $+ \infty$            \\
                          $++$                   & $ \mathrm{\textcolor{red}{DD}}_t$ & \& &$\Big(T_{i,t}(0)<T_{i,t}(1)\Big)$     &\textcolor{red}{$\Big(*,T_{i,t}(0), 1\Big)$} &$<$& \textcolor{red}{$(*,T_{i,t}(1), 1\Big)$} & $+ \infty$ \\
                          $+$                   & $ \mathrm{LL}_t$ &  \& & $\Big(Y_{i,t}(0)<Y_{i,t}( 1)\Big)$ & $\Big(Y_{i,t}(0),T_{i,t}(0), 0\Big)$  &$<$& $\Big(Y_{i,t}(1),T_{i,t}(1), 0\Big)$  & $Y_{i,t}(1)-Y_{i,t}(0)>0$   \\
                          $=$                   & $ \mathrm{\textcolor{red}{DD}}_t$  & \& & $\Big(T_{i,t}(0)=T_{i,t}( 1)\Big)$    &\textcolor{red}{$\Big(*,T_{i,t}(0), 1\Big)$} &$=$& \textcolor{red}{$\Big(*,T_{i,t}(1),1\Big)$} & $0$ \\
                          $=$                   & $ \mathrm{LL}_t$ &  \& & $\Big(Y_{i,t}(0)=Y_{i,t}(1)\Big)$   &$\Big(Y_{i,t}(0), T_{i,t}(0), 0\Big)$  &$=$& $\Big(Y_{i,t}(1),T_{i,t}(1), 0\Big)$  & $Y_{i,t}(1)-Y_{i,t}(0) =0$   \\
 $-$                   & $ \mathrm{LL}_t$ & \& & $\Big(Y_{i,t}(0)>Y_{i,t}(1)\Big)$ & $\Big(Y_{i,t}(0),T_{i,t}(0), 0\Big)$   &$>$& $\Big(Y_{i,t}(1),T_{i,t}(1), 0\Big)$  & $Y_{i,t}(1)-Y_{i,t}(0)<0$   \\
                          $--$                   & $ \mathrm{\textcolor{red}{DD}}_t$ & \& & $\Big(T_{i,t}(0)>T_{i,t}(1)\Big)$         &\textcolor{red}{$\Big(*,T_{i,t}(0),1\Big)$} &$>$& \textcolor{red}{$\Big(*,T_{i,t}(1),1\Big)$} & $- \infty$ \\
                     $---$                   & $ \mathrm{L\textcolor{red}{D}}_t$    & &      &$\Big(Y_{i,t}(0), T_{i,t}(0),0\Big)$ &$>$& \textcolor{red}{$\Big(*,T_{i,t}(1),1\Big)$}  & $- \infty$            \\ \hline
\end{tabular}
\end{center}
\caption{Composite outcome order and difference under the assumption of death as the worst outcome}
\textcolor{red}{red}: death occurred, +: beneficial effect, -: harmfull effect, $\mathrm{LL}_t$: always-survivor stratum, $\mathrm{DL}_t$: protected stratum, $\mathrm{LD}_t$: harmed stratum, $\mathrm{DD}_t$: never-survivor stratum, $*$: does not belong to the outcome space of real numbers, $W$: binary treatment, $Y_{i,t}(W)$: longitudinal potential outcome, $\Big(T_{i,t}(W), D_{i,t}(W)\Big)$: time-to-event potential outcomes
\label{tab:composit_outcome}
\end{table*}

\subsubsection{Median of the composite outcome}

Now that we have defined the difference of composite potential outcomes, we could be tempted to compute an average causal effect on the composite potential outcomes. Yet, the difference of composite potential outcomes can take infinite values, and thus the average causal effect should not be computed. However, certain percentiles of the distribution may remain finite and would provide a measure of the “magnitude of better health”. Building on this idea, the median of the composite outcome (MCO) estimand can be defined as the median of the difference of the two composite potential outcomes:
\begin{align}
   \label{eq:estimand_sim} \tau^{\mathrm{MCO}}_{\mathrm{fs}}(t) =  \left\{
        \begin{aligned}
        &[Z_{i,t}(1) - Z_{i,t}(0)]_{\frac{N + 1}{2}} &\text{$N$ odd}\\
          \frac{1}{2}\Big(&[Z_{i,t}(1) - Z_{i,t}(0)]_{\frac{N + 1}{2}} &\text{$N$ even} \\
          &+ [Z_{i,t}(1) - Z_{i,t}(0)]_{\frac{N + 1}{2}+1}\Big)& 
        \end{aligned}
        \right. 
\end{align}
For the MCO to remain finite, and thus fully interpretable, it requires determining a priori whether the median—or any chosen percentile—will be different from infinity. This estimand is an individual version of the marginal treatment effect, the survival incorporated median, introduced by Xiang et al. \cite{xiangetalSurvivalincorporatedMedianVs2023} and might lead to different conclusions (Appendix A). 

\subsubsection{Pairwise Comparison}\label{section: pairwise}
An alternative is to consider an estimand that remains well defined and finite in all circumstances. Such an estimand can be based solely on the ordering of the composite potential outcomes. Because the composite outcome represents an overall measure of health, a natural choice of estimand is the probability that the composite potential outcome under active treatment is better than the one under control treatment. Such estimand is the pairwise comparison (PC) estimand:  
\begin{align}
    \label{eq:estimand_pc1} \tau^{\mathrm{PC}}_{\mathrm{fs}}(t) = \frac{1}{N}\displaystyle\sum_{i=1}^N \Big\{&\mathds{1}\left(Z_{i,t}(1) > Z_{i,t}(0)\right) \\
   \nonumber &+ \frac{1}{2} \mathds{1}\left(Z_{i,t}(1) = Z_{i,t}(0)\right)\Big\}
\end{align}
Yet other estimands have been used, such as the Wilcoxon-Mann-Whitney (WMW) estimand. It can be defined as the probability that for a pair of randomly selected individuals (one under active treatment and one under control treatment), the composite outcome under active treatment is higher than the one under control treatment:
\begin{eqnarray*}
    \tau^{\mathrm{WMW}}_{\mathrm{fs}}(t) =& \frac{1}{\text{Card}\left(\{i, W_i = 1\}\right)}&\frac{1}{\text{Card}\left(\{j, W_j = 0\}\right)}\\
    &\times \displaystyle\sum_{i, W_i = 1} \displaystyle\sum_{j, W_j = 0} \Big\{&\mathds{1}\left(Z_{i,t}(1) > Z_{j,t}(0)\right) \\
    &&+ \frac{1}{2} \mathds{1}\left(Z_{i,t}(1) = Z_{j,t}(0)\right)\Big\}
\end{eqnarray*}
where $\mathrm{Card}(\cdot)$ denotes the cardinality of a set. Fay et al.~\cite{fay_causal_2018} have shown that this quantity is a causal estimand and that it can be interpreted as "the expected change in quantile going from control to treatment". They show that the WMW estimand differs from the PC estimand and thus may lead to opposite conclusions. A phenomenon known as Hand's paradox \cite{handComparingTwoTreatments1992}. Ensuring agreement between these estimands requires strong assumptions, such as a constant additive treatment effect \cite{fay_causal_2018}. Despite this limitation, the WMW estimand was used to define the Desirability of Outcome Ranking \cite{evansUsingOutcomesAnalyze2016}, for which an ALS-specific version was designed: Patient-Ranked Order of Function \cite{van_eijk_composite_2022}. Different variations of the WMW estimand are also known under the name of win statistics \cite{dong_win_2023}, and encompass the win ratio \cite{pocock_win_2012}, the net benefit \cite{verbeeckUnbiasednessEfficiencyNonparametric2021}, and the Combined Assessment of Function and Survival \cite{berry_combined_2013}. 
Given its clear statistical definition and causal interpretation, the PC estimand provides the most appropriate choice when constructing an estimand based on the ordering of a composite outcome and thus is favored over the WMW estimand in this article.

\subsection{Survivors Average Causal Effect}

If not trying to compare longitudinal potential outcomes beyond death, we can try to identify a subpopulation of individuals for whom the difference of longitudinal potential outcomes remains well-defined. Conditioning on an observed post-randomization variable must be avoided, as it breaks randomization and induces selection bias \cite{frangakis_principal_2002, hernan_structural_2004}. Conditioning on potential outcomes avoids this bias. To guarantee that the longitudinal potential outcomes are well defined, the analysis can be restricted to always-survivors \cite{frangakis_principal_2002}. This is the idea underlying the survivor average causal effect (SACE), defined as the mean difference between the two longitudinal potential outcomes in the principal stratum of always-survivors \cite{frangakis_principal_2002}:
\begin{align}
    \tau^{\mathrm{SACE}}_{\mathrm{fs}}(t) = \frac{1}{N_{LL_t}} \displaystyle\sum_{i: G_{i,t} = \mathrm{LL}_t} \Big\{Y_{i,t}(1) - Y_{i,t}(0)\Big\} \label{eq:estimand_sace}
\end{align}

\begin{table*}
\centering
\begin{tabular}{llcc}
\hline
Oder &Estimands & Individual-level & Finite sample \\
assumption& & $\tau_{i}(t)$ & $\tau_{\mathrm{fs}}(t)$ \\
\hline
No & ACE & $Y_{i,t}(1) - Y_{i,t}(0)$  & $\frac{1}{N} \displaystyle\sum_{i=1}^N \left\{Y_{i,t}(1) - Y_{i,t}(0)\right\}$ \\

&&&\\
No & RMST &$T_{i,u}(1) - T_{i,u}(0)$& $\frac{1}{N} \displaystyle\sum_{i=1}^N \Big\{T_{i,u}(1) - T_{i,u}(0)\Big\}$ \\
&&&\\
 No & SACE & $Y_{i,t}(1) - Y_{i,t}(0) \text{ if }  G_{i,t} = \mathrm{LL}_t$ &$\frac{1}{N_{LL}} \displaystyle\sum_{i, G_{i,t} = \mathrm{LL}_t} \Big\{Y_{i,t}(1) - Y_{i,t}(0)\Big\}$\\
 &&&\\
 Yes & PC & $\mathds{1}\left(Z_{i,t}(1) > Z_{i,t}(0)\right)$
 & $\frac{1}{N}\displaystyle\sum_i^N \Big\{\mathds{1}\left(Z_{i,t}(1) > Z_{i,t}(0)\right)$ \\

  &  &  $+ \frac{1}{2} \mathds{1}\left(Z_{i,t}(1) = Z_{i,t}(0)\right)$ &  $+ \frac{1}{2} \mathds{1}\left(Z_{i,t}(1) = Z_{i,t}(0)\right)\Big\}$\\

&&&\\
Yes &  MCO & $Z_{i,t}(1) - Z_{i,t}(0)$& $\left\{
        \begin{aligned}
        &[Z_{i,t}(1) - Z_{i,t}(0)]_{\frac{n + 1}{2}} &&\text{ for $N$ odd}\\
          &\frac{1}{2}\Big([Z_{i,t}(1) - Z_{i,t}(0)]_{\frac{n + 1}{2}} &&\text{ for $N$ even}\\
          &+ [Z_{i,t}(1) - Z_{i,t}(0)]_{\frac{n + 1}{2}+1}\Big) 
        \end{aligned}
        \right.$ \\

\hline
    \end{tabular}
    \caption{Summary of the causal estimands\\
    ACE: average causal effect, RMST: restricted mean potential survival time, SACE: survivor average causal effect, PC: pairwise comparison, MCO: median of the composite outcome, $W$: binary treatment, $Y_{i,t}(W)$: longitudinal potential outcome, $\Big(T_{i,t}(W), D_{i,t}(W)\Big)$: time-to-event potential outcomes, $Z_{i,t}(W)$: composite potential outcome
    }
    \label{tab:estimands}
\end{table*}

\section{Estimation} \label{sec:estimation}

For the estimation, we move from the potential outcomes notation ($Y_{i,t}(1), Y_{i,t}(0)$), to the observed/missing notation ($Y^{obs}_{i,t}, Y^{mis}_{i,t}$), closest to the data. Using this notation, the individual-level causal effect can be rewritten as $Y_{i,t}(1) - Y_{i,t}(0) = (2W_i-1)*(Y^{obs}_{i,t} - Y^{mis}_{i,t})$ \cite{imbens_causal_2015}. The implementation is publicly available in the repository Truncation By Death Estimators \footnote{\url{https://github.com/JulietteOrtholand/TBDEstimators}}, together with a simple illustrative example.

\subsection{Missing outcome estimators}\label{sec:estimator_outcome}

For each missing outcome, we examine identification assumptions and specify Bayesian estimators. We first describe the estimator for the time-to-event missing outcome. Then, we use it to estimate the probability of belonging to the always-survivors stratum and describe the estimator for the missing longitudinal outcome. Finally, we combine the estimators for the missing time-to-event and longitudinal outcomes to create an estimator for the missing composite outcome.

\subsubsection{Estimator for missing time-to-event outcome}\label{death_assumptions}
There exists a direct relation between the survival (Equation \ref{eq:survival_function}) and the hazard (Equation \ref{eq:hazard_function}) that can be used to model survival:
\begin{equation}
    S_{i,t}(W) = \exp\Big(\int_0^t \Lambda_{i,u}(W) \text{d}u\Big) \label{eq:rel_hazard}
\end{equation}
To limit the assumption on hazard functional form, we model it with a Bayesian piecewise-constant proportional hazards model \cite{rodrigurz_lecture_2007, citekey} :
\begin{align*}
    \nonumber \Lambda_{i,t}(W) &= \lambda_{0,t}^{u} \exp(\alpha^{u\top} X_i) \\
    \lambda_{0,t}^{u} &= \lambda^{u,j} \text{ for } t\in[\tau_{j-1},\tau_j]
\end{align*}
with $u \in \{c,a\}$, $c$ for control treatment and $a$ for active treatment, $\alpha^{u}$ the slope of covariates and $\lambda^{u,j}$ the baseline hazards between $\tau_{j-1}$ and $\tau_j$. Normal priors are assigned to the covariates' slopes, allowing for both positive and negative effects, whereas Gamma priors are assigned to the baseline hazards to ensure their positivity:
\begin{align*}
    \alpha^{u} &\sim \mathcal{N}\left(\mu_{\alpha^{u}}, \sigma_{\alpha^{u}}\right) &&&
    \lambda^{u,j} &\sim \Gamma\left(\mu_{\lambda^{u,j}}, \sigma_{\lambda^{u,j}}\right)
\end{align*}
This model is equivalent to a Poisson regression formulation applied over discrete time intervals \cite{laird_covariance_1981, holford_analysis_1980}. Specifically, let $d_{i,j}$ denote a discretized event indicator equal to 1 if individual $i$ dies in the interval $(\tau_{j-1}, \tau_j]$ and 0 otherwise. The discretized event potential outcomes thus follow a Poisson distribution:
\begin{align*}
\left.
\begin{pmatrix}
d_{i,j}(0) \\
d_{i,j}(1)
\end{pmatrix}
\,\right|\, \theta_t^S
&\sim
\operatorname{Poisson}
\begin{pmatrix}
\lambda^{c,j}\,\exp\!\bigl(\alpha^{c\top} X_i\bigr)\,(\tau_j-\tau_{j-1}) \\
\lambda^{a,j}\,\exp\!\bigl(\alpha^{a\top} X_i\bigr)\,(\tau_j-\tau_{j-1})
\end{pmatrix}
\end{align*}
with $\theta_t^S=\Bigl(\lambda^{c}, \alpha^{c},\lambda^{a}, \alpha^{a}\Bigr)$.

We make several assumptions to identify the missing discretized event outcome. First, we assume for all $i$ a \textit{stable unit treatment value assumption (SUTVA)}: no interference between individuals and a well-defined version of each treatment \cite{imbens_causal_2015}: 
\begin{eqnarray*}
    d^{obs}_{i,j} = d_{i,j}(1)\times W_i + d_{i,j}(0)\times(1-W_i)
\end{eqnarray*}
We also assume an \textit{individualistic assignment} \cite{imbens_causal_2015}: treatment is assigned at an individual level. 
Finally, we assume a \textit{strong ignorable treatment assignment} \cite{rosenbaumCentralRolePropensity1983}, composed of two aspects:
\begin{itemize}
    \item \textit{Unconfounded Assignment:} given covariates, treatment allocation is as good as randomly assigned, 
    \begin{eqnarray*}
        \Pr(W|X,d_{j}(1), d_{j}(0)) = \Pr(W|X)
    \end{eqnarray*}
    for all $W$, $X$, $d_{j}(1)$, $d_{j}(0)$. 
    \item \textit{Probabilistic Assignment:} each individual $i$ has a strictly positive probability of receiving either treatment,
    \begin{eqnarray*}
        0 < p_i(X,d_{j}(1), d_{j}(0)) < 1
    \end{eqnarray*}
\end{itemize}
These assumptions are likely to hold in an RCT. 

Under these assumptions, the missing discretized indicator for $t\in[\tau_{j-1},\tau_j]$ can be written as depending on observed quantities:
\begin{eqnarray*}
    \delta_t(W_i, X_i) = \Big(&&W_i\lambda^{a,j} \exp(\alpha^{a\top} X_i) \\
    &&+ (1 - W_i)\lambda^{c,j} \exp(\alpha^{c\top} X_i)\Big)
    \times(\tau_j - \tau_{j-1})
\end{eqnarray*}
\begin{eqnarray*}
    d^{\mathrm{mis}}_{i,j}|d^{\mathrm{obs}}_{i,j},W_i, X_i,\theta^S \sim &&\operatorname{Poisson}\left(\delta_t(W_i, X_i)\right)\\
    d^{\mathrm{obs}}_{i,j}|W_i, X_i,\theta^S \sim &&\operatorname{Poisson}\left(\delta_t(W_i, X_i)\right)
\end{eqnarray*}
The posterior distribution of such an estimator admits no closed form. The estimation is thus carried out via the Hamiltonian Monte Carlo algorithm with a No-U-Turn sampler \cite{hoffmanNoUTurnSamplerAdaptivelya} available in the \texttt{pymc} library \cite{pymc2023, pymc_example}. 

%

\subsubsection{Estimator for missing longitudinal outcome}\label{long_assumptions}

We model longitudinal potential outcomes of always-survivors as linear functions of baseline covariates. Since only one potential outcome is observed for each individual, the data contain no empirical information on the correlation between the two potential outcomes \cite{imbens_causal_2015}. To avoid introducing unsupported dependence in the imputation process, we assume that the two potential outcomes are uncorrelated:
\begin{align*}
\left.
\begin{matrix}
Y_{i,t}(0) \\
Y_{i,t}(1)
\end{matrix}
\,\right| \theta_t^Y, G_{i,t} = \mathrm{LL}_t
&\sim
\mathcal{N}
\left(
\begin{pmatrix}
\beta_t^{c,0} + X_i^\top \beta_t^{c,1} \\
\beta_t^{a,0} + X_i^\top \beta_t^{a,1}
\end{pmatrix},
\begin{pmatrix}
\sigma_t & 0 \\
0 & \sigma_t
\end{pmatrix}
\right)
\end{align*}
with $c$ for control treatment, $a$ for active treatment, $(\beta_t^{c,0},\, \beta_t^{a,0})$ the intercepts, $(\beta_t^{c,1},\, \beta_t^{a,1})$ the slope vectors, $\sigma_t$ the standard deviation of the residual variability, and $\theta_t^Y=\Bigl(\beta_t^{c,0},\, \beta_t^{a,0},\, \beta_t^{c,1},\, \beta_t^{a,1}\Bigr)$. We use Half-Normal distribution prior ($\mathcal{N}^+$) to guarantee the positivity of the standard deviation of the residual variability, and Normal distribution priors for the rest of the parameters:
\begin{align*}
\sigma_t\sim\mathcal{N}^+\!\left(0,\, \sigma_{\sigma_t}\right) &&& \beta_t^{u,k}\sim \mathcal{N}\!\left(\mu_{\beta_t^{u,k}},\sigma_{\beta_t^{u,k}}\right)
\end{align*}
with $k \in \{0,1\}$ denoting the intercept or the slope, and $u \in \{c,a\}$ control and active treatment. 

To identify the missing potential longitudinal outcomes, we assume \textit{SUTVA} and \textit{individualistic assignment} for the longitudinal potential outcome (see Section \ref{death_assumptions} for more details).
%
As in Grossi et al. \cite{grossiBayesianPrincipalStratification2025}, we additionally assume a \textit{strong ignorable treatment assignment}  \cite{rosenbaumCentralRolePropensity1983}:
\begin{itemize}
    \item \textit{Unconfounded Assignment:} given covariates, treatment allocation is as good as randomly assigned, 
    \begin{eqnarray*}
        \Pr(W|X, G_{t}, Y_{t}(1), Y_{t}(0)) = \Pr(W|X)
    \end{eqnarray*}
    for all $W$, $X$, $Y_{t}(1)$, $Y_{t}(0)$, $G_{t}$
    \item \textit{Probabilistic Assignment:} each individual $i$ has a strictly positive probability of receiving either treatment
    \begin{equation*}
        0 < p_i(X, G_{t}, Y_{t}(1), Y_{t}(0)) < 1
    \end{equation*}
\end{itemize}
Unconfounded treatment assignment for the longitudinal potential outcome requires a stronger assumption than that needed for the identification of time-to-event outcomes due to the condition on the always-survivors stratum. Specifically, it requires both substantive knowledge of the disease process and the availability of appropriate measured covariates to identify the stratum \cite{frangakis_principal_2002}. To identify the stratum, we assume \textit{SUTVA} and \textit{individualistic assignment} for the time-to-event potential outcomes and a \textit{strong ignorable assignment} (see section \ref{death_assumptions}). 

Following Mattei et al. \cite{matteiAssessingCausalEffects2025}, we specify two models: one for membership in the always-survivor principal stratum  (conditionally on the observed covariates), and another for the distribution of the potential longitudinal outcomes (conditionally on always-survivor status and covariates).
For the principal strata model, the probability that a given individual is an always-survivor at time $t$ is denoted $h_{i,t}$ and defined as the product of survival under both treatments:
\begin{align*}
    h_{i,t} = \Pr(G_{i,t} = \mathrm{LL}_t| X_i) = S^{\mathrm{mis}}_{i,t}\times S^{\mathrm{obs}}_{i,t}
\end{align*}
We use the estimator described in Section~\ref{death_assumptions} to estimate the missing survivals. To approximate the posterior distribution, we draw parameter values $\theta^{S,j}_t$, indexed by $j$, from the posterior and work with the corresponding scalar quantities. We then construct an estimator for the probability of being an always-survivor as the product of the estimated missing survival and a boolean equal to one if the individual is observed alive:
\begin{align*}
    \widehat{h}^{j}_{i,t} = \widehat{S}^{\mathrm{mis},j}_{i,t} \times \mathds{1}\left((D^{\mathrm{obs}}_{i,t} =0) \text{ \& }(T^{\mathrm{obs}}_{i,t}>t)\right)
\end{align*}
The linear model for the missing longitudinal outcome of always-survivors can then be written conditioning on observed quantities:
\begin{align*}
    &\gamma_t (W_i, X_i)= W_i(\beta^{a,0}_t + X_i^T\beta^{a,1}_t) + (1 - W_i)(\beta^{c,0}_t + X_i^T\beta^{c,1}_t) \\
    &Y^{\mathrm{mis}}_{i,t}|Y^{\mathrm{obs}}_{i,t},(G_{i,t}^j = \mathrm{LL}_t), W_i, X_i, \theta^Y_t \sim \mathcal{N}\left(\gamma_t (W_i, X_i), \sigma_t \right)\\
    &Y^{\mathrm{obs}}_{i,t}|(G_{i,t}^j = \mathrm{LL}_t), W_i, X_i, \theta^Y_t \sim \mathcal{N}\left(\gamma_t (W_i, X_i), \sigma_t \right)
\end{align*}
We repeatedly sample the always-survivor status under a Bernoulli distribution with a parameter equal to the probability of being an always-survivor. Thus, the likelihood of the longitudinal model directly depends on this probability (more details are provided in Appendix B):
\begin{eqnarray*}
    &&\log \left(\Pr(Y^{\mathrm{obs}}_{t}| (\widehat{G}_{t}^j = \mathrm{LL}_t), W, X, \theta^Y_t)\right) \\
    &&= \displaystyle\sum_{i=1}^N \widehat{h}^j_{i,t}\Big( - \log\left(\sigma_t \sqrt{2\pi}\right) - \frac{1}{2\sigma_t^2} \Big( \gamma_t (W_i, X_i) - Y^{\mathrm{obs}}_{i,t}\Big)^2\Big)
\end{eqnarray*}
The posterior distribution of this estimator admits no closed form. The estimation is thus carried out via the Hamiltonian Monte Carlo algorithm with a No-U-Turn sampler \cite{hoffmanNoUTurnSamplerAdaptivelya}, available in the \texttt{pymc} library \cite{pymc2023, pymc_example}. Readers familiar with the joint modeling literature may recognize similarities with the likelihood used in joint models for longitudinal and time-to-event data \cite{rizopoulos_joint_2012}. However, it is important to emphasize that our approach is fundamentally a two-step procedure: the survival model is first estimated, and the resulting survival probabilities are then used to weight the longitudinal likelihood. This ordering reflects the causal structure of the problem, as survival determines whether the longitudinal outcome is a real number, but not vice versa. 

\subsubsection{Estimator for missing composite outcome}\label{section:estimator_metric}

The composite outcome is a three-dimensional vector comprising both longitudinal and time-to-event components. Therefore, the two estimators developed in Sections \ref{long_assumptions} and \ref{death_assumptions} can be combined to impute the missing composite outcome. The difference of composite potential outcomes is central to the estimation of estimands using composite outcomes. We describe here how it is estimated, building on its definition summarized in Table \ref{tab:composit_outcome}. 

First, we can consider the case where the individual $i$ is observed alive by time $t$ (first condition of Equation \ref{eq:est_composit}). If the individual is alive under the missing scenario (always-survivor stratum), the difference of composite potential outcomes coincides with the difference of longitudinal potential outcomes (Equation \ref{eq:comp_long_equal}). If the individual is dead under the missing scenario (protected or harmed strata), the difference of composite potential outcomes is set to: $+\infty$ if the individual is observed under active treatment (protected stratum); and $-\infty$ if the individual is observed under control treatment (harmed stratum). The individual's status under the missing scenario is uncertain; thus the individual is considered alive with a probability equal to the missing survival. The difference of composite potential outcomes thus takes two values with different probabilities. 

Second, if the individual $i$ is observed dead by time $t$ (second condition of Equation \ref{eq:est_composit}). Here, the difference of composite potential outcomes is always equal to infinity, and only the sign varies depending on whether i) the missing death occurs before or after the observed death, ii) the observed treatment. If the missing death occurs after the observed death, the difference of composite potential outcomes is: positive under observed control treatment, and negative under observed active treatment; both with a probability equal to the survival at time of observed death. If the missing death occurs before the observed death, the difference of composite potential outcomes is negative under observed control treatment, and positive under observed active treatment, both with a probability equal to one minus the survival at time of observed death. Again, the difference of composite potential outcomes takes the two values with the different probabilities.
\begin{eqnarray}
    &(2W_i - 1) (Z^{\mathrm{obs}}_{i,t} - \widehat{Z}^{\mathrm{mis}}_{i,t})= \label{eq:est_composit}\\
    &\nonumber \left\{
        \begin{aligned}
        &\text{ if } T^{\mathrm{obs}}_{i,t} > t \\
        &\left\{
        \begin{aligned}
        & (2W_i - 1) (Y^{\mathrm{obs}}_{i,t} - \widehat{Y}^{\mathrm{mis}}_{i,t}) \text{, with } \Pr(T^{\mathrm{mis}}_{i,t} > t) = \widehat{S}^{\mathrm{mis}}_{i,t}\\
        & (2W_i - 1) (+\infty) \text{, with } \Pr(T^{\mathrm{mis}}_{i,t} \leq t) = 1 - \widehat{S}^{\mathrm{mis}}_{i,t}
        \end{aligned}
        \right.\\
        &\text{ if } T^{\mathrm{obs}}_{i,t} \leq t \\
        &\left\{
        \begin{aligned}
        & (2W_i - 1) (-\infty) \text{, with } \Pr(T^{\mathrm{mis}}_{i,t} > T^{\mathrm{obs}}_{i,t}) = \widehat{S}^{\mathrm{mis}}_{i,T^{\mathrm{obs}}_{i,t}} \\
        & (2W_i - 1)  (+\infty) \text{, with } \Pr(T^{\mathrm{mis}}_{i,t} \leq T^{\mathrm{obs}}_{i,t} ) = 1 - \widehat{S}^{\mathrm{mis}}_{i,T^{\mathrm{obs}}_{i,t}}
        \end{aligned}
        \right.
        \end{aligned}
        \right.
\end{eqnarray}

\subsection{Treatment effect estimators}\label{treatment_estimators}

Estimators for the different treatment effects rely on missing outcome models (Section \ref{sec:estimator_outcome}). We first present estimators that use composite outcomes, then the remaining. 

\subsubsection{Estimators with comparison assumption}

To estimate causal estimands built upon composite potential outcomes, the main challenge stems from estimating the difference of composite potential outcomes (Section \ref{section:estimator_metric}). 
To estimate the MCO estimand (Equation~\ref{eq:estimand_sim}), we use the values of the difference of composite potential outcomes with their probabilities, to define a weighted median estimator that accounts for uncertainty of stratum membership:
\begin{align*}
   \widehat{\tau}^{\mathrm{MCO}}_{\mathrm{fs}}(t) =  \left\{
        \begin{aligned}
        &[(2W_i-1)\times (Z^{\mathrm{obs}}_{i,t} - \widehat{Z}^{\mathrm{mis}}_{i,t})]_{\frac{N + 1}{2}}  \text{ $N$ odd}\\
          \frac{1}{2}\Big(&[(2W_i-1)\times (Z^{\mathrm{obs}}_{i,t} - \widehat{Z}^{\mathrm{mis}}_{i,t})]_{\frac{N + 1}{2}} \text{ $N$ even}\\
          &+ [(2W_i-1)\times (Z^{\mathrm{obs}}_{i,t} - \widehat{Z}^{\mathrm{mis}}_{i,t})]_{\frac{N + 1}{2}+1}\Big)
        \end{aligned}
        \right.
\end{align*}
To estimate the PC estimand (Equation~\ref{eq:estimand_pc1}), we use the average over individuals, of the probability for the difference of composite potential outcomes to be superior or equal to zero:
\begin{align*}
\widehat{\tau}^{PC}_{fs}(t) = \frac{1}{N} \displaystyle\sum_{i=1}^N \Big\{&\Pr\left((2W_i-1)(Z^{\mathrm{obs}}_{i,t}- \widehat{Z}^{\mathrm{mis}}_{i,t})>0\right) \\
&+\frac{\Pr\left((2W_i-1)(Z^{\mathrm{obs}}_{i,t}- \widehat{Z}^{\mathrm{mis}}_{i,t})=0\right)}{2}\Big\}
\end{align*}

\subsubsection{Estimators without order assumption}

The RMST estimand, $\tau^{\mathrm{RMST}}_{\mathrm{sp}}(t)$ (Equation~\ref{eq:estimand_rmst}) is a linear function of the time-to-event potential outcomes and therefore presents no particular estimation challenges. We estimate it using the average difference between the observed time to death and the integral of the missing survival:
\begin{align*}
     \widehat{\tau}^{RMST}_{fs}(t) =& \displaystyle\sum_{i=1} ^N \left\{(2W_i-1) \left(T^{\mathrm{obs}}_{i,t} - \int_0^t \widehat{S}^{\mathrm{mis}}_{i,u}du\right) \right\}
\end{align*}
%
The principal challenge in estimating the SACE estimand (Equation~\ref{eq:estimand_sace}) lies in the identification of the always-survivors. We address this problem by weighting the difference longitudinal potential outcomes by the probability of being an always-survivor:
\begin{align*}
\widehat{\tau}^{SACE}_{fs}(t) = \frac{1}{\displaystyle\sum_{i=1}^N \widehat{h}^{j}_{i,t}} 
\displaystyle\sum_{i=1}^N \left\{\widehat{h}^{j}_{i,t} (2W_i-1)
 (Y^{\mathrm{obs}}_{i,t} - \widehat{Y}^{\mathrm{mis}}_{i,t})\right\}
\end{align*}
%

\section{Simulation study}\label{sec:simulation}

The simulation study follows the Aims, Data-generating mechanisms, Methods, Estimands, Performance measures framework \cite{morrisUsingSimulationStudies2019a}. Its aim is to assess the validity of the estimators and to examine their behavior across a range of scenarios. The experiments can be reproduced with publicly available code \footnote{\url{https://github.com/JulietteOrtholand/truncation_by_death/tree/main}}.

\subsection{Method}\label{sect:sim_method}

For each individual $i$, we consider a single baseline covariate $X_i$. The longitudinal outcome is simulated under a linear progression model, and times-to-event outcomes follow an accelerated failure time model with a Weibull distribution ($\mathcal{W}$):
\begin{eqnarray}
    &Y_{i,t}(W) \sim \mathcal{N}(f(W, X_i), \sigma)\label{eq:long_sim}\\
    \nonumber &f(W, X_i) = ((a^0_0  + a^0_X X_{i} ) + W_i \times(a^1_0 + a^1_X X_{i}))*t \\
    &T_{i,\infty}(W) \sim \mathcal{W}\left( g(W, X_i), \rho \right) \label{eq:surv_sim}\\
    \nonumber &g(W, X_i) = (\theta^0_0  + \theta^0_X X_{i}) + W \times(\theta^1_0 + \theta^1_X X_{i}) 
\end{eqnarray}
We simulate an RCT with a follow-up time of 15 months and measurement times $t = 3, 6, 9, 12, 15$.
Four scenarios are considered: i) no treatment effect and no censoring (named No effect, No censoring); ii) no treatment effect with censoring (named No effect); iii) a beneficial treatment effect on both time-to-event and the longitudinal outcomes (named Beneficial); and iv) opposing treatment effects on time-to-event (harmful) and the longitudinal outcome (beneficial) (named Mixed). Under each scenario, 100 datasets are simulated. The parameter values used for simulation are summarized in Supplementary A.

For each scenario and each individual, we estimate the missing potential outcomes at time $t \in \{3, 6, 9, 12, 15\}$. One hundred samples are drawn from the posterior distribution of the survival model to estimate always-survivors. Model estimation is performed using identical prior distributions under control and active treatments ($u\in \{c, a\}$), thereby encoding the absence of prior information on treatment effects. Weakly informative priors are specified for covariate effects in the survival model ($\alpha^{u} \sim \mathcal{N}(0,1)$) and in the longitudinal model ($\beta^{u,1}_t \sim \mathcal{N}(0,100)$). The prior for the residual variability of the longitudinal model is also defined as weakly informative: $\sigma_t \sim \mathcal{N}^+(0,100)$. We assume prior knowledge on the longitudinal slope, and the prior is set to the simulated value: $\beta^{u,0}_t \sim \mathcal{N}(-2,3)$.
The simulated median survivals range between 18 and 22 months (Supplementary A). Assuming a constant hazard over time and calibrating it to yield median survivals of 18 and 22 months (Equation \ref{eq:rel_hazard}), we obtain  $\lambda^{u,j} \in [0.032, 0.038]$. We specify the priors for the baseline hazard accordingly: $\lambda^{u,j} \sim \Gamma(0.035, 0.1)$. Additional experiments with less well-calibrated priors are reported in Appendix C.
We then compute the finite-sample estimators defined in Section \ref{treatment_estimators} for RMST, SACE, PC, and MOC. 
Performance is assessed in terms of bias and coverage using the estimands computed on simulated potential outcomes \cite{morrisUsingSimulationStudies2019a}.

\subsection{Results}

Posterior medians and associated credibility intervals of the treatment effect estimators for each scenario are displayed in 
Figure~\ref{fig:validation_sim}. 
Estimators exhibited good results on coverage and bias, which are reported in Supplementary A. \\
The RMST estimator performance improves as the number of observed deaths increases (Figure~\ref{fig:validation_sim}). 
In contrast, the SACE, the MCO, and the PC estimators exhibit increasing uncertainty as the number of deaths rises, reflected in wider credibility intervals (Figure~\ref{fig:validation_sim}).\\
%
The MCO estimate takes infinite values at later time points in the beneficial and mixed scenarios (Figure \ref{fig:validation_sim}). This is due to a high number of deaths in the protected stratum (beneficial scenario) and in the harmed stratum (mixed scenario).
%
The PC estimator is particularly sensitive to the prior specification of no treatment effect (Figure \ref{fig:validation_sim}). Still, despite the priors assuming no treatment effect, the PC estimator indicates a positive treatment effect in the beneficial scenario. \\
%
In the mixed scenario (last column of Figure~\ref{fig:validation_sim}), the PC estimator reflects the evolving balance between the effects on longitudinal and time-to-event outcomes: at early time points, when few deaths have occurred, the estimator exceeds 0.5, indicating a beneficial effect on the longitudinal outcome; as mortality increases under active treatment, the estimator decreases below 0.5, indicating a harmful effect. In the mixed scenario, when used alone, the SACE and the MCO only retrieve the beneficial impact of the treatment on the longitudinal outcome. A complete characterization of the mixed scenario therefore requires considering them jointly with RMST.

\begin{figure}
    \centering
    \includegraphics[width=\columnwidth]{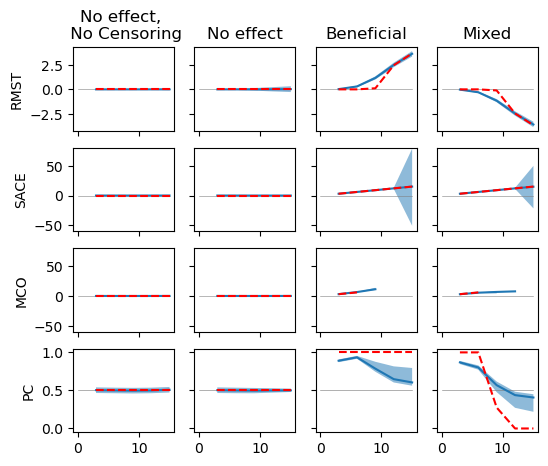}
    \caption{Estimated treatment effects on the simulated scenarios \\
    \underline{Legend:} RMST: restricted mean potential survival time, SACE: survivor average causal effect, PC: pairwise comparison, MCO: median of the composite outcome. The grey line represents no treatment effect, the red line represents the simulated median treatment effect. The blue plain line with clear area represents the estimated median with centered 95\% credibility intervals.}
    \label{fig:validation_sim}
\end{figure}

\section{Application}\label{sec:application}

\subsection{Method}

We analyze data from a double-blind RCT in individuals with ALS. The RCT was run to assess the efficacy of Olesoxime on 512 individuals \cite{lengletetalPhaseIIIII2014}. The trial accepted the null hypothesis on time-to-event outcomes, based on the log-rank test. An improvement in ALSFRS-R score at 9 months (not sustained at 18 months) was estimated using an inappropriate estimate computed on observed survivors \cite{frangakis_principal_2002, hernan_structural_2004}.\\
The longitudinal outcome is the ALSFRS-R score change from baseline and the time-to-event outcome is the time to death from baseline.  We are interested in the treatment effect at 1, 3, 6, 12, and 18 months. Time-to-event outcome identification relies on \textit{SUTVA} and an \textit{individual, strong ignorable treatment assignment} (section \ref{death_assumptions}), which are reasonable assumptions in RCTs. Longitudinal and composite outcome identification relies on \textit{SUTVA} and an \textit{individual strong ignorable treatment assignment} (section \ref{long_assumptions}). The latter assumption is assumed given the following covariates: ALSFRS-R at baseline, sex, site of symptom onset, age at first symptom, and delay between first symptoms and baseline \cite{grassanoSexDifferencesAmyotrophic}. We assume that, conditionally on the observed covariates, additional missingness is at random. 

We use the estimators described in Section \ref{sec:estimator_outcome} for the missing potential outcomes and the treatment effects. One hundred samples are drawn from the posterior distribution of the survival model to estimate always-survivors. We model ALSFRS-R score change with a normal distribution parametrised by a linear model, thus approximating the discrete score as i) a continuous variable ii) with no floor or ceiling effect \cite{gordonProgressionALSNot2010}. These approximations are reasonable as i) the ALSFRS-R score change is quite granular over 18 months, and ii) individuals were recruited with a baseline ALSFRS-R score that limits the floor and ceiling effect of the score ranging from 0 to 48 (control: 38.2 (5.2), active: 39.1 (4.8)) \cite{benatarPrognosticClinicalBiological2024}. As in the simulation study, we specify identical priors under active and control treatments, assuming no treatment effect. We use weakly informative priors for covariate effects and for the residual variability of the longitudinal model (see Section~\ref{sect:sim_method}). ALSFRS-R change is approximately 0.92 units per month in RCTs, so we specified priors for the longitudinal slope accordingly: $\beta^{u,0}_t \sim \mathcal{N}(-1, 3)$ \cite{castrillo-vigueraClinicalSignificanceChange2010}. 
Individuals typically die 3 to 4 years after symptom onset \cite{talbott_epidemiology_2016}, and were enrolled on average 1.5 years after onset (Appendix D). This corresponds to an expected survival of 18 to 30 months from trial baseline. Assuming a constant hazard over time integrated to 18 and 30 months to get a median survival (Equation \ref{eq:rel_hazard}), we specify baseline hazard priors as $\lambda^{u,j} \sim \Gamma(0.035, 0.1)$. The additional missing values are imputed using the missing outcome estimators: first for survival and subsequently for the longitudinal outcome.\\

\subsection{Results} 

Baseline characteristics were similar between treatment arms with respect to site of onset, sex, baseline ALSFRS-R score, age at symptom onset, and time from symptom onset to baseline (Appendix D). Overall, 24\% of individuals had died by the end of follow-up (Appendix D). 
Figure~\ref{fig:trophos_results} summarizes the estimated treatment effects also available in Appendix D. For both the RMST and SACE estimands, the posterior medians suggest a beneficial treatment effect, although the corresponding 95\% credibility intervals include the null. Consistent with this pattern, the MCO and PC estimands yield more pronounced effects, particularly at 12 months, with 95\% credibility intervals of $[0.53, 0.60]$ for the PC and $[0.78, 2.54]$ for the MCO.

\begin{figure*}
    \centering
    \includegraphics[width=2\columnwidth]{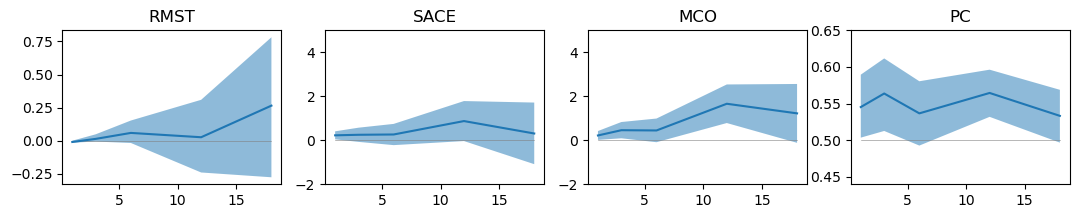}
    \caption{Estimated treatment effects of Olesoxime on RCT data \\
    \underline{Legend:} RMST: restricted mean potential survival time, SACE: survivor average causal effect, PC: pairwise comparison, MCO: median of the composite outcome. The grey line represents no treatment effect. The blue plain line with clear area represents the estimated median with centered 95\% credibility intervals.}
    \label{fig:trophos_results}
\end{figure*}
\section{Discussion}
%
In this article, we reviewed and compared estimands defined with longitudinal outcomes truncated by death. We then illustrated their behavior on both simulated and real RCT data to support their practical use.

When individuals die, their longitudinal outcomes no longer belong to the outcome space of real numbers \cite{frangakis_principal_2002}. We show that three types of estimands exist, among which two require an assumption to compare longitudinal outcomes after death: hypothetical and composite estimands. Hypothetical estimands require that death can be intervened on without affecting the longitudinal outcome. Yet, the same biological process often drives both longitudinal and time-to-event outcomes, making this assumption highly implausible. The composite estimands require assuming an order between the longitudinal outcome and death. Yet, such an ordering assumption is not intuitive and should be left to individual appreciation, especially in chronic diseases. Finally, the survivor average causal effect does not require such type of assumptions by computing the treatment effect only on individuals that would have survived under both treatments. Thus, it seems the most suited estimand to deal with truncation by death on chronic diseases like ALS.  
%
Identification of the different causal estimands using the longitudinal outcome relies on the same assumptions: \textit{SUTVA and individual strong ignorable treatment assignment}. Assuming the unconfoundingness of the assignment of the longitudinal outcome is a strong assumption and requires substantive knowledge of the disease and the availability of appropriate measured covariates.

We compared the behavior of the treatment effect estimators in a simulation study using Bayesian estimators. Among composite estimands, the results confirm that the pairwise comparison estimand captures the trade-off between time-to-event and longitudinal outcomes when treatment affects both time-to-event and the longitudinal outcomes in opposite directions. We highlight that the median of the composite outcome can take infinite values, limiting its practicality. When associated with restricted mean survival time, survivor average causal effect provides a more complete but inherently multidimensional characterization. We have pointed to the persistent influence of prior specifications, as discussed for survivor average causal effect by Richardson et al.~\cite{richardsonTransparentParametrizationsModels2011}. 

We illustrated this problem using data from an RCT in ALS, providing a concrete example of the practical application of the estimators. A \textit{strong ignorable treatment assignment assumption} was assumed, given available covariates: ALSFRS-R at baseline, sex, site of symptom onset, age at first symptom, and delay between first symptoms and baseline \cite{grassanoSexDifferencesAmyotrophic}. Nevertheless, other factors are known to play a role, such as genetic factors \cite{fGeneticsSexPathogenesis2020}. We also observed that when the treatment has a beneficial effect on both time-to-event and longitudinal outcomes, the estimated effect tends to be more pronounced for estimands based on composite outcomes at the cost of requiring ordering assumptions.

We compared estimands for longitudinal data truncated by death from a theoretical and practical point of view. The
code for the Bayesian estimators is available on GitHub\footnote{\url{https://github.com/JulietteOrtholand/TBDEstimators}}, together with a simple example to facilitate their use.
Yet implementation relies on relatively simple parametric models, which could be extended in future work. 

To conclude, this work aims to support robust analyses of RCTs in life-threatening diseases. We argue that, in the presence of death truncation, pairing the survivor average causal effect with the restricted mean survival time estimand provides an interpretable characterization of treatment effects on longitudinal and time-to-event outcomes.

\begin{appendix}

\section{Median Composite Outcome estimand and Survival Incorporated Median} \label{sec:MCOSIM}

We consider the hypothetical population described in Table \ref{tab:toy_ex}. In such a population, the Median Composite Outcome estimand \cite{xiangetalSurvivalincorporatedMedianVs2023} is equal to -1:
\begin{align*}
    &M\Big[&&(0-3), (0-2), (4-6), (4-5), \\
    &&&(*-*), (4-4),  (4-1)) \Big]\\
    =& M\Big[&&-3, -2, -2, -1, (*-*),  0, 3\Big] = -1
\end{align*}
with $M$ the median. Whereas the Survival Incorporated Median is equal to 1:
\begin{align*}
    &M(*,0,0,4,4,4,4) 
    - M(*,1,2,3,4,5,6) \\
    &= 4 - 3 = 1
\end{align*}
This illustrates that the two estimands correspond to different quantities, a consequence of the median nonlinearity. 
\begin{table*}
    \centering
    \begin{tabular}{cccccccccc}
\hline
Individuals & Time & \multicolumn{2}{c}{Longitudinal outcome}& \multicolumn{2}{c}{Indicator of death} & \multicolumn{2}{c}{Time of death}\\
 $i$ & $t$ (months) & $Y_{i,t}(0)$ & $Y_{i,t}(1)$ & $D_{i,t}(0)$ & $D_{i,t}(1)$ & $T_{i,t}(0)$ & $T_{i,t}(1)$ \\
\hline
1 & 18 &  1 & 4 & 0 & 0 & 18 & 18 \\
2 & 18 &  2 & 0 & 0 & 0 & 18 & 18 \\
3 & 18 &  3 & 0 &  0 & 0 & 18 & 18 \\
4 & 18 &  4 & 4 &  0 & 0 & 18 & 18 \\
5 & 18 &  5 & 4 &  0 & 0 & 18 & 18 \\
6 & 18 &  6 & 4 &  0 & 0 & 18 & 18 \\
7 & 18 &  * & * &  1 & 1 & 8 & 8 \\
\hline
    \end{tabular}
    \caption{Example of population on which Median Composite Outcome and Survival Incorporated Median estimand lead to opposite conclusions}
    \label{tab:toy_ex}
    $*$: does not belong to the outcome space of real numbers 
\end{table*}

\section{Missing Longitudinal potential outcome likelihood}\label{app:likelihood}
The estimated always-survivor status is represented by the Boolean $\widehat{G}^k_{i,t} = LL_t$. When drawing it over $K$ draws for each individual from a Bernoulli distribution of parameter $\widehat{h}_{i,t}$, it follows that:
\begin{eqnarray*}
    & \frac{1}{K}\displaystyle\sum_{k=0}^K\displaystyle\sum_{i=1}^N\log \left(\Pr(Y^{\mathrm{obs}}_{i,t}| (\widehat{G}^k_{i,t} = \mathrm{LL}_t), W_i, X_i, \theta^Y_t)\right) \\
    &= \frac{1}{K}\displaystyle\sum_{k=0}^K\displaystyle\sum_{i=1}^N 1(\widehat{G}^k_{i,t} = \mathrm{LL}_t)\Big( - \log\Big(\sigma_t \sqrt{2\pi}\Big) \\
    & - \frac{1}{2\sigma_t^2} \left( \gamma_t(W_i, X_i) - Y^{\mathrm{obs}}_{i,t}\right)^2\Big) \\
    &= \displaystyle\sum_{i=1}^N \left[\frac{1}{K}\displaystyle\sum_{k=0}^K 1(\widehat{G}^k_{i,t} = \mathrm{LL}_t)\right]\Big( - \log\left(\sigma_t \sqrt{2\pi}\right) \\
    &- \frac{1}{2\sigma_t^2} \left( \gamma_t(W_i, X_i) - Y^{\mathrm{obs}}_{i,t}\right)^2\Big)
\end{eqnarray*}
where $\frac{1}{K}\displaystyle\sum_{k=0}^K1(\widehat{G}^k_{i,t} = \mathrm{LL}_t)$ is an estimator of $\widehat{h}_{i,t}$. 

\section{Prior sensitivity}\label{appendix:prior_sensitivity}

\begin{figure}
    \centering
    \includegraphics[width=\columnwidth]{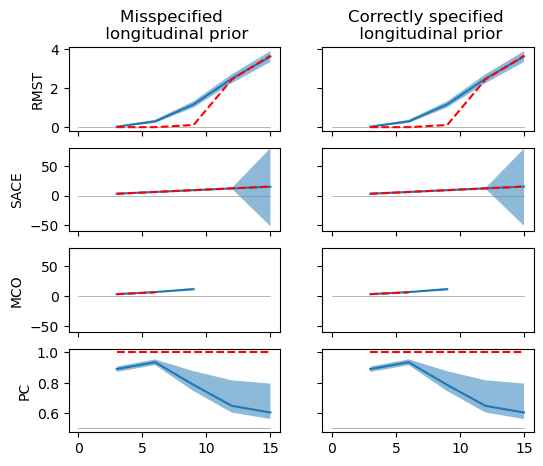}
    \caption{Estimated treatment effects for the beneficial scenarios depending on prior\\
\underline{Legend:} RMST: restricted mean potential survival time, SACE: survivor average causal effect, PC: pairwise comparison, MCO: median of the composite outcome. The grey line represents no treatment effect. The grey line represents no treatment effect, and the red line represents the simulated median treatment effect. The blue plain line with clear area represents the estimated median with centred 95\% credibility intervals.}
    \label{fig:prior_impact}
\end{figure}

We compare results under a correctly specified prior for the longitudinal slope, $\beta^{u,0}_t \sim \mathcal{N}(-2,3)$, and a misspecified prior, $\beta^{u,0}_t \sim \mathcal{N}(-1,3)$, in the beneficial scenario. Figure~\ref{fig:prior_impact} indicates no difference between the corresponding estimates.

\section{Application}

\begin{figure}
    \centering
    \includegraphics[width=\columnwidth]{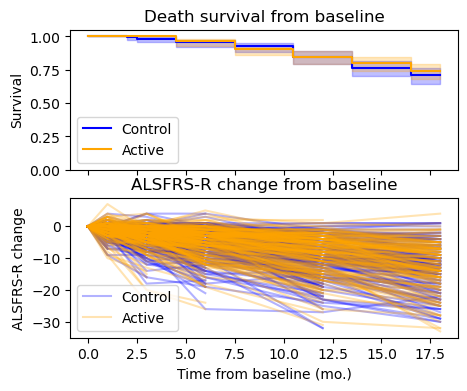}
    \caption{Survival and ALSFRS-R change from baseline on individuals from the Olesoxime RCT}
    \label{fig:description_trophos}
\end{figure}

Characteristics of the RCT cohort are presented in Table \ref{tab:trophos_stats}. Figure \ref{fig:description_trophos} displays longitudinal trajectories of individuals. Table \ref{tab:res_trophos} contains the estimated treatment effects at 1, 3, 6, 12 and 18 months. 

\begin{table}
\begin{center}
\resizebox{\columnwidth}{!}{
\begin{tabular}{lrrr}
\hline
 & Control& Active& p-value\\
\hline
Individuals& 252& 258& - \\
Visits& 1,274& 1,321& - \\
 Site of symptoms onset (\% of bulbar) & 19.8\%& 19.4\%& - \\
Sex (\% of female) & 35.3\%& 35.3\%& - \\
Baseline ALSFRS-R & 38.2 (5.2)& 39.1 (4.8)& 0.5\\
 Age at first symptoms (years) & 54.3 (11.2)& 56.0 (11.2)& 0.1\\
 Delay to baseline (years)& 1.4 (0.9)& 1.3 (0.8)& 0.2\\
 \hline
\end{tabular}
}
\end{center}
\caption{Baseline characteristics of individuals from the Olesoxime RCT}
\label{tab:trophos_stats}
\end{table}

\begin{table}
\begin{center}
\resizebox{\columnwidth}{!}{
\begin{tabular}{lrrrrr}
\hline
 Time& \multicolumn{4}{c}{Treatment effects} & Death  \\
 (mo.)& RMST & SACE & MCO & PC & (\%) \\
\hline
1 & [-0.01, -0.0] & [0.04, 0.41] & [0.02, 0.42] & [0.50, 0.59] & 0.00 \\
3 & [-0.01, 0.05] & [-0.07, 0.57] & [0.09, 0.83] & [0.51, 0.61] & 0.79 \\
6 & [-0.02, 0.15] & [-0.22, 0.74] & [-0.09, 0.99] & [0.49, 0.58] & 3.93 \\
12 & [-0.24, 0.31] & [-0.04, 1.78] & [0.78, 2.54] & [0.53, 0.6] & 13.95 \\
18 & [-0.28, 0.78] & [-1.09, 1.71] & [-0.12, 2.56] & [0.50, 0.57] & 23.97 \\
\hline
\end{tabular}
}
\end{center}
\caption{Estimates of the treatment effects of Olesoxime}
mo.: months, RMST: restricted mean potential survival time, SACE: survivor average causal effect, PC: pairwise comparison, MCO: median of the composite outcome. 
\label{tab:res_trophos}
\end{table}

\end{appendix}

\newpage
\begin{funding}
The first author was supported by the "Investissements d’avenir" by the French government under the management of Agence Nationale de la Recherche (reference ANR-10-IAIHU-06, ANR-19-P3IA-0001 (PRAIRIE 3IA Institute), ANR-19-JPW2-000 (E-DADS)).
\end{funding}

\begin{supplement}
\stitle{Supplementary A}
\sdescription{It contains the values of parameters used for the simulation study, and coverages and biases of the estimators.}
\end{supplement}


\bibliographystyle{imsart-number} 
\bibliography{bibliography}       
\clearpage
\onecolumn
\section*{Supplementary A: Simulation main study}
Data were simulated using parameters in Table \ref{tab:param_summary}.  Coverages of the estimators are available in Table \ref{tab:coverage_validation}, and bias in Table \ref{tab:bias_validation}.
\begin{table}[b]
\begin{center}
\begin{tabular}{lrrrrrrrrrrrr}
\hline
&\multicolumn{6}{c}{Longitudinal parameters} & \multicolumn{6}{c}{Survival parameters}\\
 & $a^0_0$& $a^1_0$ & $a^0_x$ & $a^1_x$ & $a^0_u$ & $a^1_u$ & $\theta^0_0$& $\theta^1_0$ & $\theta^0_x$ & $\theta^1_x$ & $\theta^0_u$ & $\theta^1_u$ \\
\hline
no effect, no censoring & -2 & -2 & 1 & 1 & 0 & 0 & 150 &150 & 1 & 1 & 0 & 0 \\
 no effect & -2 & -2 & 1 & 1 & 0 & 0 & 25 & 25 & 1 & 1 & 0 & 0 \\
 beneficial & -2 & -1 & 1 & 1 & 0 & 0 & 25 & 20 & 1 & 1 & 0 & 0 \\
 mixed & -2 & -1 & 1 & 1 & 0 & 0 & 25 & 20 & 1 & 1 & 0 & 0 \\
\hline
\end{tabular}
\end{center}
\caption{Parameters used to simulate data scenarios.\\
Data is simulated with $\sigma = 0.05*48$, $\rho = 3.5$}
\label{tab:param_summary}
\end{table}
\begin{table}[b]
\begin{center}
\begin{tabular}{llrrrrrrrrr}
\hline
Scenario & Time & \multicolumn{2}{c}{SACE} & \multicolumn{2}{c}{PC} & \multicolumn{2}{c}{MCO} &  \multicolumn{2}{c}{RMST} & Death  \\
 & & sim. & coverage & sim. & coverage & sim. & coverage &  sim. & coverage & (\%)  \\
\hline
no censure &3 & 0.00 & 68.00 & 0.50 & 68.00 & 0.00 & 68.00 & 0.00 & 26.00 & [0.0, 0.0] \\
no effect &6 & 0.00 & 67.00 & 0.50 & 67.00 & 0.00 & 67.00 & 0.00 & 45.00 & [0.0, 0.0] \\
&9 & 0.00 & 69.00 & 0.50 & 72.00 & 0.00 & 72.00 & 0.00 & 71.00 & [0.0, 0.0] \\
&12 & 0.00 & 72.00 & 0.50 & 72.00 & 0.00 & 72.00 & 0.00 & 89.00 & [0.0, 0.0] \\
&15 & 0.00 & 65.00 & 0.50 & 71.00 & 0.00 & 71.00 & 0.00 & 98.00 & [0.0, 0.0] \\
\hline
no effect &3 & 0.00 & 67.00 & 0.50 & 65.00 & 0.00 & 65.00 & 0.00 & 37.00 & [0.0, 1.0] \\
 &6 & 0.00 & 67.00 & 0.50 & 61.00 & 0.00 & 61.00 & 0.00 & 65.00 & [2.0, 6.0] \\
&9 & 0.00 & 80.00 & 0.50 & 80.00 & 0.00 & 80.00 & 0.00 & 78.00 & [13.0, 19.0] \\
&12 & 0.00 & 84.00 & 0.50 & 77.00 & 0.00 & 77.00 & 0.00 & 77.00 & [33.0, 42.0] \\
&15 & 0.00 & 98.00 & 0.50 & 83.00 & 0.00 & 81.00 & 0.00 & 73.00 & [59.0, 68.0] \\
\hline
beneficial &3 & 3.00 & 68.00 & 1.00 & 0.00 & 3.00 & 61.00 & 0.01 & 29.00 & [0.0, 2.0] \\
&6 & 6.00 & 70.00 & 1.00 & 0.00 & 6.00 & 14.00 & 0.17 & 22.00 & [8.0, 13.0] \\
&9 & 9.00 & 83.00 & 1.00 & 0.00 & - & - & 0.87 & 14.00 & [30.0, 37.0] \\
&12 & 12.00 & 99.00 & 1.00 & 0.00 & - & - & 2.16 & 31.00 & [57.0, 64.0] \\
&15 & 15.00 & 97.00 & 1.00 & 0.00 & - & - & 3.42 & 65.00 & [77.0, 84.0] \\
\hline
mixed&3 & 3.00 & 70.00 & 0.99 & 0.00 & 3.00 & 66.00 & -0.01 & 43.00 & [0.0, 2.0] \\
&6 & 6.00 & 65.00 & 0.84 & 27.00 & 6.00 & 22.00 & -0.17 & 25.00 & [7.0, 13.0] \\
&9 & 9.00 & 83.00 & 0.49 & 74.00 & - & - & -0.87 & 16.00 & [30.0, 37.0] \\
&12 & 12.00 & 97.00 & 0.16 & 0.00 & - & - & -2.16 & 31.00 & [57.0, 64.0] \\
&15 & 15.00 & 86.00 & 0.03 & 0.00 & - & - & -3.42 & 64.00 & [78.0, 84.0] \\
\hline
\end{tabular}
\end{center}
\caption{Coverages for the validation scenarios \\
\underline{Legend:} RMST: restricted mean potential survival time, SACE: survivor average causal effect, PC: pairwise comparison, MCO: median of the composite outcome. The grey line represents no treatment effect. sim: real value simulated on the data, 100 datasets simulated for each scenario, "-" are present for the MCO treatment effect where the simulated treatment effect was infinite.}
\label{tab:coverage_validation}
\end{table}

\begin{table*}
\begin{center}
\begin{tabular}{llrrrrrrrrr}
\hline
Scenario & Time & \multicolumn{2}{c}{SACE} & \multicolumn{2}{c}{PC} & \multicolumn{2}{c}{MCO} &  \multicolumn{2}{c}{RMST} & Death  \\
 & & sim. & bias & sim. & bias & sim. & bias &  sim. & bias & (\%)  \\
\hline
no censure  &3 & 0.00 & [-0.2, 0.24] & 0.50 & [-0.03, 0.04] & 0.00 & [-0.2, 0.26] & 0.00 & [-0.01, 0.01] & [0.0, 0.0] \\
no effect &6 & 0.00 & [-0.22, 0.21] & 0.50 & [-0.04, 0.04] & 0.00 & [-0.22, 0.23] & 0.00 & [-0.03, 0.02] & [0.0, 0.0] \\
&9 & 0.00 & [-0.23, 0.2] & 0.50 & [-0.04, 0.03] & 0.00 & [-0.24, 0.21] & 0.00 & [-0.04, 0.04] & [0.0, 0.0] \\
&12 & 0.00 & [-0.21, 0.21] & 0.50 & [-0.04, 0.04] & 0.00 & [-0.22, 0.23] & 0.00 & [-0.06, 0.06] & [0.0, 0.0] \\
&15 & 0.00 & [-0.16, 0.26] & 0.50 & [-0.03, 0.05] & 0.00 & [-0.17, 0.28] & 0.00 & [-0.08, 0.08] & [0.0, 0.0] \\
\hline
no effect &3 & 0.00 & [-0.2, 0.24] & 0.50 & [-0.03, 0.04] & 0.00 & [-0.21, 0.26] & 0.00 & [-0.02, 0.02] & [0.0, 1.0] \\
 &6 & 0.00 & [-0.23, 0.23] & 0.50 & [-0.04, 0.04] & 0.00 & [-0.25, 0.26] & 0.00 & [-0.06, 0.06] & [2.0, 6.0] \\
&9 & 0.00 & [-0.29, 0.24] & 0.50 & [-0.04, 0.03] & 0.00 & [-0.35, 0.31] & 0.00 & [-0.13, 0.13] & [13.0, 19.0] \\
&12 & 0.00 & [-0.41, 0.37] & 0.50 & [-0.03, 0.03] & 0.00 & [-0.55, 0.56] & 0.00 & [-0.21, 0.22] & [33.0, 42.0] \\
&15 & 0.00 & [-2.46, 2.34] & 0.50 & [-0.02, 0.02] & 0.00 & [-3.91, 4.32] & 0.00 & [-0.28, 0.32] & [59.0, 68.0] \\
\hline
beneficial &3 & 3.00 & [-0.22, 0.23] & 1.00 & [-0.14, -0.1] & 3.00 & [-0.16, 0.32] & 0.01 & [-0.0, 0.04] & [0.0, 2.0] \\
&6 & 6.00 & [-0.25, 0.26] & 1.00 & [-0.08, -0.05] & 6.00 & [0.26, 0.91] & 0.17 & [0.05, 0.21] & [8.0, 13.0] \\
&9 & 9.00 & [-0.42, 0.4] & 1.00 & [-0.25, -0.12] & - & - & 0.87 & [0.13, 0.45] & [30.0, 37.0] \\
&12 & 12.00 & [-4.13, 2.75] & 1.00 & [-0.4, -0.19] & - & - & 2.16 & [0.09, 0.55] & [57.0, 64.0] \\
&15 & 15.00 & [-18.59, 9.07] & 1.00 & [-0.44, -0.21] & - & - & 3.42 & [-0.08, 0.48] & [77.0, 84.0] \\
\hline
mixed &3 & 3.00 & [-0.22, 0.22] & 0.99 & [-0.15, -0.11] & 3.00 & [-0.3, 0.18] & -0.01 & [-0.05, 0.0] & [0.0, 2.0] \\
&6 & 6.00 & [-0.27, 0.24] & 0.84 & [-0.07, -0.01] & 6.00 & [-0.92, -0.26] & -0.17 & [-0.21, -0.04] & [7.0, 13.0] \\
&9 & 9.00 & [-0.51, 0.3] & 0.49 & [-0.01, 0.13] & - & - & -0.87 & [-0.44, -0.12] & [30.0, 37.0] \\
&12 & 12.00 & [-4.04, 2.65] & 0.16 & [0.1, 0.31] & - & - & -2.16 & [-0.52, -0.06] & [57.0, 64.0] \\
&15 & 15.00 & [-20.78, 5.87] & 0.03 & [0.19, 0.42] & - & - & -3.42 & [-0.43, 0.12] & [78.0, 84.0] \\
\hline
\end{tabular}

\end{center}
\caption{Bias for the validation scenarios \\
\underline{Legend:}RMST: restricted mean potential survival time, SACE: survivor average causal effect, PC: pairwise comparison, MCO: median of the composite outcome. The grey line represents no treatment effect. sim: real value simulated on the data, 100 datasets simulated for each scenario, "-" are present for the MCO treatment effect where the simulated treatment effect was infinite}
\label{tab:bias_validation}
\end{table*}

\end{document}